\documentclass[aps, prl, letterpaper, english, reprint, tightenlines, citeautoscript, groupedaddress, superscriptaddress, 10pt]{revtex4-1}

\usepackage[T1]{fontenc}
\usepackage[latin9]{inputenc}
\usepackage{babel}
\usepackage{textcomp}
\usepackage{amssymb, amsmath}
\usepackage{graphicx}
\usepackage{bm}
\usepackage{tikz}
\usepackage[colorlinks=true, linkcolor=black, citecolor=blue, urlcolor=blue, unicode=true]{hyperref}


\begin{document}

\newcommand{\cso}{Cu$_{2}$OSeO$_{3}$}
\newcommand{\suppl}{Supplementary Materials \cite{supplement}}

\title{Non-Reciprocal Zone Boundary Magnon Propagation in \cso{}}

\newcommand{\ill}{Institut Laue-Langevin (ILL), 71 avenue des Martyrs, 38000 Grenoble, France}
\newcommand{\tum}{Physik-Department, Technische Universit\"at M\"unchen (TUM), James-Franck-Str. 1, 85748 Garching, Germany}
\newcommand{\mlz}{Heinz-Maier-Leibnitz-Zentrum (MLZ), Technische Universit\"at M\"unchen (TUM), Lichtenbergstr. 1, 85747 Garching, Germany}
\newcommand{\frm}{Forschungsneutronenquelle Heinz-Maier-Leibnitz (FRM-II), Lichtenbergstr. 1, 85747 Garching, Germany}
\newcommand{\epfl}{\'Ecole Polytechnique F\'ed\'erale de Lausanne (EPFL), CH-1015 Lausanne, Switzerland}
\newcommand{\fkf}{Max-Planck-Institut f\"ur Festk\"orperforschung, Heisenbergstr. 1, D-70569 Stuttgart, Germany}
\newcommand{\jcns}{J\"ulich Centre for Neutron Science (JCNS), Lichtenbergstr. 1, 85748 Garching, Germany}
\newcommand{\mcqst}{Munich Center for Quantum Science and Technology (MCQST), Schellingstr. 4, 80799 Munich, Germany}
\newcommand{\zqe}{Zentrum f\"ur QuantumEngineering (ZQE), Am Coulombwall 3a, 85748 Garching, Germany}

\author{T. Weber}
\email[Corresponding author: ]{tobias.weber@ill.fr}
\affiliation{\ill}

\author{N. Heinsdorf}
\affiliation{\fkf}

\author{M. Stekiel}
\affiliation{\tum}
\affiliation{\jcns}

\author{P. Steffens}
\affiliation{\ill}


\author{A. P. Schnyder}
\affiliation{\fkf}

\author{C. Pfleiderer}
\affiliation{\tum}
\affiliation{\frm}
\affiliation{\mcqst}
\affiliation{\zqe}

\date{\today}

\begin{abstract}
Inelastic neutron scattering in the chiral magnet \cso{} reveals strong non-reciprocal effects on magnon
propagation at the boundary of the nuclear Brillouin zone.
The non-reciprocal response is strongest at a central position between the zone corner and edge mid-point.
We explain these results using an effective linear spin-wave model.
While directional effects in chiral magnets have so far only been known to exist at low momenta
close to the center of the Brillouin zone,
the present study shows that non-reciprocity persists at the highest possible reduced momenta.
The observed magnons show very little damping within the limits of our experimental resolution,
making them of great interest for the fundamental research on compact,
high-frequency magnonic applications.
\end{abstract}

\maketitle

Non-reciprocity or directional dichroism has attracted considerable interest in recent years
as the effect has been observed for
photons \cite{Sounas2017, Zhang2018, Jia2025},
phonons \cite{Xu2019, Shan2023, Ren2025},
and magnons \cite{Grigoriev2015, Sato2016, Seki2016, cheong2018, Seki2020, Che2024, Weber2022skx}.
Non-reciprocal (NR) magnon propagation is of great importance in the creation of directional
magnonic devices such as spin-wave diodes \cite{Lan2015, Szulc2020, Chen2022},
antennas with energy flow in one direction \cite{Devolder2023},
unidirectional magnonic couplers \cite{Grachev2024},
wave guides that are selective to certain frequencies or programmable magnonic logic gates \cite{Odintsov2022}.
Materials with low damping and therefore a long lifetime of the spin-waves are of special importance to
magnonic applications \cite{Flebus2024, Han2024}.

Studies reported to date focused on very low reduced momenta close to the center of the first nuclear Brillouin zone, $\Gamma$.
With the ongoing efforts to develop high-speed, compact antiferromagnetic magnonics, spin-waves at large
momenta are of great interest \cite{Schoenfeld2025}.
This also raises the question if NR magnons persist at large momenta up to the zone boundary.
Such observations would open new possibilities for the tunability of the modes and the development
of antiferromagnetic spintronic and magnonic devices operating coherently at much higher frequencies than for low momenta.
While microwave spectroscopy of antiferromagnetic magnons is possible in principle,
exploratory work to clarify the possible existence of high-momentum NR magnons
is best addressed using inelastic neutron scattering, which is not restricted to zero reduced momentum, $q=0$.
Although a previous search for NR zone boundary magnons yielded no conclusive results \cite{Wilde2021},
a very recent study could identify such an effect \cite{Quanchao2025}.

NR excitations of energy $E$ and reduced momentum $q$ propagate unidirectionally in a crystal.
They have no corresponding mode moving in the opposite direction at $-q$ \cite{Sato2019},
where $q = Q - G$, with total momentum $Q$ and lattice vector $G$.
In its weakest form, non-reciprocity manifests itself in a difference of spectral weights for magnon
creation versus annihilation while keeping the same dispersion $E \left(q \right)$.
For the dynamical structure factor $S \left(q, E \right)$ this means that
$S \left(q, -E \right) \neq \exp \left(-E/k_B T \right) S \left(q, E \right)$ \cite{Squires1996}.
In a more fundamental form, the shape of the dispersion itself changes,
$E \left(q \right) \neq E \left(-q \right)$.
In this form, excitations of a reduced momentum $q$ that are created with an
energy $E$ are not annihilated at $-E$, but at a different magnitude of energy than for creation.

In magnetic systems, non-reciprocity is inherent to non-centrosymmetric compounds where the lack of spatial
inversion symmetry enables a Dzyaloshinskii-Moriya interaction (DMI).
The DMI adds skew-symmetric contributions to the magnetic interaction matrix $J$
which leads to a shift between the positive and the negative energy eigenvalues of the system's Hamiltonian \cite{Toth2015},
whose absolute values would otherwise be equal without the DMI.
While spatial inversion is always broken in NR systems,
the additionally broken time-reversal symmetry that is usually encountered in NR magnon dispersions
is not a strict requirement.
Generally, NR effects have also been observed in systems with no broken time-reversal symmetry,
for example in the circular dichroism of light polarization \cite{Tokura2018}.

The insulator \cso{} and the itinerant magnet MnSi are model compounds to study magnons in chiral systems \cite{Schwarze2015}.
Both compounds crystallize in space group $\mathrm{P2_13}$ and share a qualitatively similar magnetic phase diagram.
Starting from the helimagnetic phase, \cso{} becomes ferrimagnetic above a second critical
field of $B_{c2}^{[001]} \approx 50\ \mathrm{mT}$  for fields along $[001]$
and $B_{c2}^{\left[110 \right]} \approx 65\ \mathrm{mT}$ for fields along $\left[110\right]$ at
$10\ \mathrm{K}$ \cite{Qian2018},
respectively, with a magnetic moment of 0.53 $\mu_B$ per Cu ion \cite{Belesi2011}.
For MnSi the boundary to field-polarized ferromagnetic order is at $B_{c2} \approx 600\ \mathrm{mT}$
for $10\ \mathrm{K}$ \cite{Bauer2012}.

Pioneering studies on NR dynamics in chiral magnets were performed by
Grigoriev \textit{et al.} \cite{Grigoriev2015} and Sato \textit{et al.} \cite{Sato2016},
who discovered that at low $q$  the dispersion branch of the field-polarized phase in MnSi
is a parabola that is off-centered from the $\Gamma$ point of the Brillouin zone.
First studies on NR dynamics in \cso{} were reported by
Seki \textit{et al.} \cite{Seki2016} and Grigoriev \textit{et al.} \cite{Grigoriev2019}.
Follow-up studies discovered that the non-reciprocity in the conical phase of both \cso{} \cite{Ogawa2021}  
and MnSi \cite{Weber2018coni, Weber2019coni} manifests itself in an asymmetric distribution
of the spectral weights between parabolic dispersion branches centered around the incommensurate
magnetic satellite Bragg peaks.
In the conical phase, the dispersion forms a band structure perpendicular to the helical
propagation vector \cite{Janoschek2010, Kugler2015}.
Magnetic excitations in the skyrmion phase of both \cso{} \cite{Che2024, Seki2020} and MnSi \cite{Weber2022skx}
take the form of energetically closely spaced Landau levels which show NR characteristics for
excitations along the skyrmion tubes.

Full mappings of the magnetic dynamics in \cso{} at momenta as high as the boundary
of the nuclear Brillouin zone were performed by Portnichenko \textit{et al.} \cite{Portnichenko2016},
Tucker \textit{et al.} \cite{Tucker2016}, and Luo \textit{et al.} \cite{luo2020low}.
Very recent studies investigated the temperature-dependence of the spin-wave stiffness \cite{Ukleev2025}.
Zhang \textit{et al.} \cite{zhang2020magnonic} found Weyl band crossing points near the zone center $\Gamma$
and the corner $R$.
However, all studies were performed without the application of an external field,
remained in the helical phase, and did not allow to discriminate NR effects.

The present study explores directional phenomena in the high-$q$ modes in \cso{} and
identifies a strong NR effect of the magnons at the zone boundary.
The magnons show no appreciable damping.
As insulator, \cso{} is ideally suited for probing high-$q$ modes, whereas
in the metallic MnSi, the Stoner continuum generates spin damping and
severely restricts the possible momenta of well-defined magnons to very low values
in the order of several helimagnetic Brillouin zones \cite{Ishikawa76, Ishikawa77}.

We measured the dynamical structure factor of field-polarized, ferrimagnetic \cso{} in a large
single-crystal of  $m \approx 5\ \mathrm{g}$.
Experiments were performed at the cold triple-axis spectrometer \textit{Thales} \cite{Thales},
where a fixed outgoing neutron wavenumber of $k_f = 1.5\, \textup{\AA}^{-1}$ was selected.
The experiments were done using open collimation, focusing Si(111) monochromator
and focusing HOPG(002) analyzer crystal arrays.
In order to filter out higher-order neutron scattering, a cooled beryllium filter was placed in the
outgoing $k_f$ axis of the spectrometer.

The sample crystal was oriented in the $\left(hhl \right)$ plane.
A horizontal magnetic field of $B = 0.4 \ \mathrm{T}$ was provided using an \textit{Oxford} superconducting cryomagnet \cite{MagnetH}.
Before each series of scans we changed the direction of the magnetic field outside the field-polarized ferrimagnetic phase
at nearly $80\ \mathrm{K}$, which is well above the magnetic ordering temperature $T_c \approx 60\ \mathrm{K}$ \cite{Qian2018}, and field-cooled the sample.
Measuring in the field-polarized phase ensures a defined spin order,
making non-reciprocity detectable by either flipping the sign of the field or the reduced momentum, since
$E\left(-q, B \right) = E\left(q, -B \right)$,
where the first option is preferred experimentally (see the \suppl{}).

Due to its very high flux both at cold and at thermal energies, the triple-axis spectrometer \textit{Thales}
is capable of selecting the necessary high energy transfers of around 10 meV, which
-- at the employed fixed $k_f = 1.5\, \textup{\AA}^{-1}$ -- corresponds to a
thermal wavenumber of $k_i = 2.66\, \textup{\AA}^{-1}$ for the incoming neutron beam.
This makes the instrument uniquely suited for providing the necessary resolution,
while also having enough neutron intensity.

For the measurements the sample was cooled to $T = 10 \ \mathrm{K}$.
Because of the low final energy at \textit{Thales} in the cold-neutron range,
the measurements could be performed around a low-$Q$ Bragg peak.
We chose the $G = \left(220\right)$ reflection for its
strong magnetic form factor as well as to suppress any phonon intensity that would interfere at larger $G$.

\begin{figure*}[htb]
\begin{centering}
\begin{tikzpicture}
	\draw (0, 0) node [ inner sep = 0 ] { \includegraphics[width=0.3\textwidth]{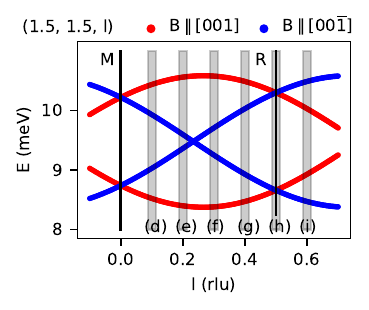} };
	\draw (-2.35, 1.5) node { \bf \fontfamily{phv} \selectfont (a) };
\end{tikzpicture}
\begin{tikzpicture}
	\draw (0, 0) node [ inner sep = 0 ] { \includegraphics[width=0.3\textwidth]{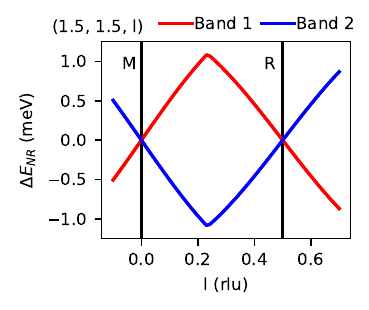} };
	\draw (-2.35, 1.5) node { \bf \fontfamily{phv} \selectfont (b) };
\end{tikzpicture}
\begin{tikzpicture}
	\draw (0, 0) node [ inner sep = 0 ] { \includegraphics[width=0.3\textwidth]{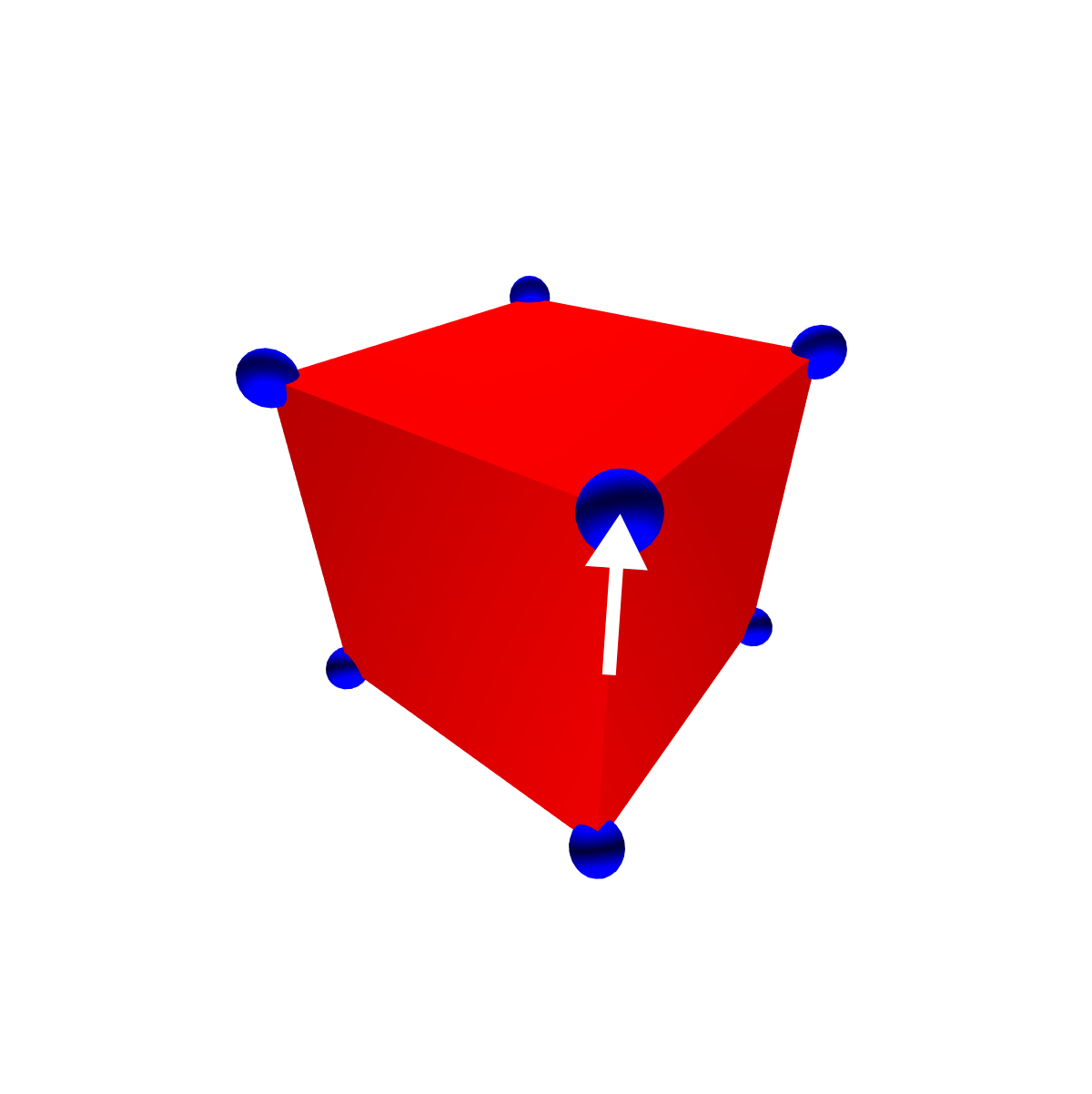} };
	\draw (-1.6, -1.5) node [ inner sep = 0 ] { \includegraphics[width=0.08\textwidth]{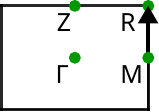} };
	\draw (-2.35, 1.1) node { \bf \fontfamily{phv} \selectfont (c) };
	\draw (0., 0.) node { \bf \fontfamily{phv} \selectfont R };
	\draw (0.05, -0.6) node { \bf \fontfamily{phv} \selectfont M };
	\draw (0., 0.8) node { \bf \fontfamily{phv} \selectfont Z };
\end{tikzpicture}
\begin{tikzpicture}
	\draw (0, 0) node [ inner sep = 0 ] { \includegraphics[width=0.3\textwidth]{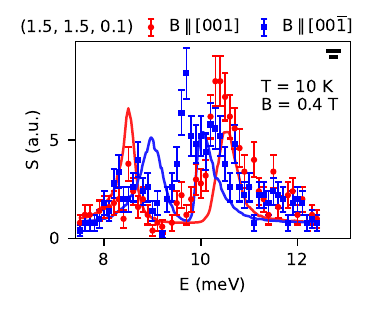} };
	\draw (-2.35, 1.5) node { \bf \fontfamily{phv} \selectfont (d) };
\end{tikzpicture}
\begin{tikzpicture}
	\draw (0, 0) node [ inner sep = 0 ] { \includegraphics[width=0.3\textwidth]{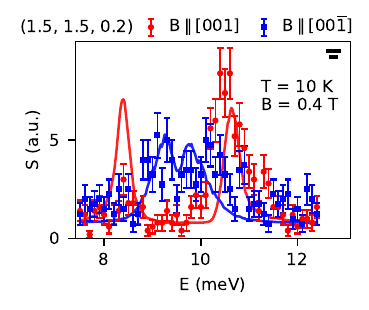} };
	\draw (-2.35, 1.5) node { \bf \fontfamily{phv} \selectfont (e) };
\end{tikzpicture}
\begin{tikzpicture}
	\draw (0, 0) node [ inner sep = 0 ] { \includegraphics[width=0.3\textwidth]{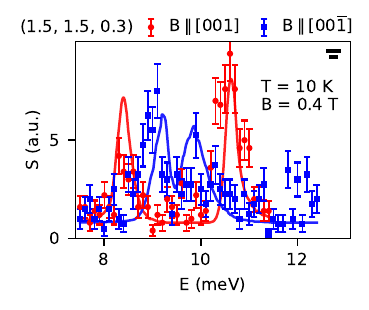} };
	\draw (-2.35, 1.5) node { \bf \fontfamily{phv} \selectfont (f) };
\end{tikzpicture}
\begin{tikzpicture}
	\draw (0, 0) node [ inner sep = 0 ] { \includegraphics[width=0.3\textwidth]{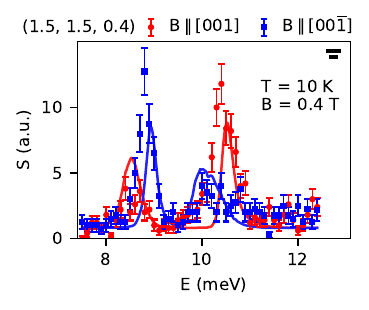} };
	\draw (-2.35, 1.5) node { \bf \fontfamily{phv} \selectfont (g) };
\end{tikzpicture}
\begin{tikzpicture}
	\draw (0, 0) node [ inner sep = 0 ] { \includegraphics[width=0.3\textwidth]{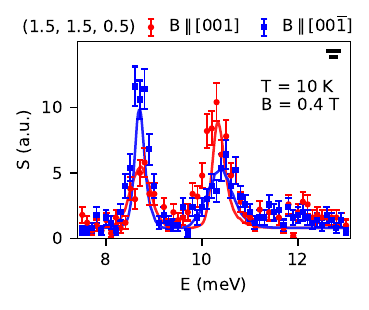} };
	\draw (-2.35, 1.5) node { \bf \fontfamily{phv} \selectfont (h) };
\end{tikzpicture}
\begin{tikzpicture}
	\draw (0, 0) node [ inner sep = 0 ] { \includegraphics[width=0.3\textwidth]{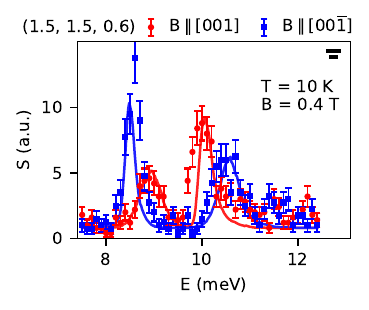} };
	\draw (-2.35, 1.5) node { \bf \fontfamily{phv} \selectfont (i) };
\end{tikzpicture}
	\caption{
	Non-reciprocal zone-boundary magnon dispersion.
	(a) Theoretical dispersion relation along the cubic zone boundary at
	$\mathbf{Q} = \left(1.5, \  1.5, \  l \right)$
	between the $M$ and the $R$ point for two field directions that are $180^{\circ}$ apart.
	The thickness of the lines signifies the spin-spin correlation function,
	which gives the spectral weight of a magnon mode.
	The positions of the individual scans are marked as gray vertical bars.
	(b) Energy shifts $\Delta E_{NR} = \left|E \left(q \right)\right| - \left|E \left(-q \right)\right|$ of the magnon bands due to non-reciprocity.
	(c) Cubic Brillouin zone, $M$, $R$, and $Z$ symmetry points and accessible zone cut within the scattering plane (inset).
	The scan path is marked with an arrow.
	(d)-(i) Data measured at \textit{Thales} (points),
	where the typical measurement time per point was about two minutes.
	The field directions $\mathbf{B} \parallel \left[ 00\bar{1} \right]$ and $\mathbf{B} \parallel \left[ 001 \right]$
	are shown in blue and red, respectively.
	The solid lines show the resolution-convolution fit of the theory and the data.
	The black lines in the top-right corners denote the incoherent and the coherent resolution at $E = 10\ \mathrm{meV}$, respectively.}
	\label{fig:B001}
\end{centering}
\end{figure*}
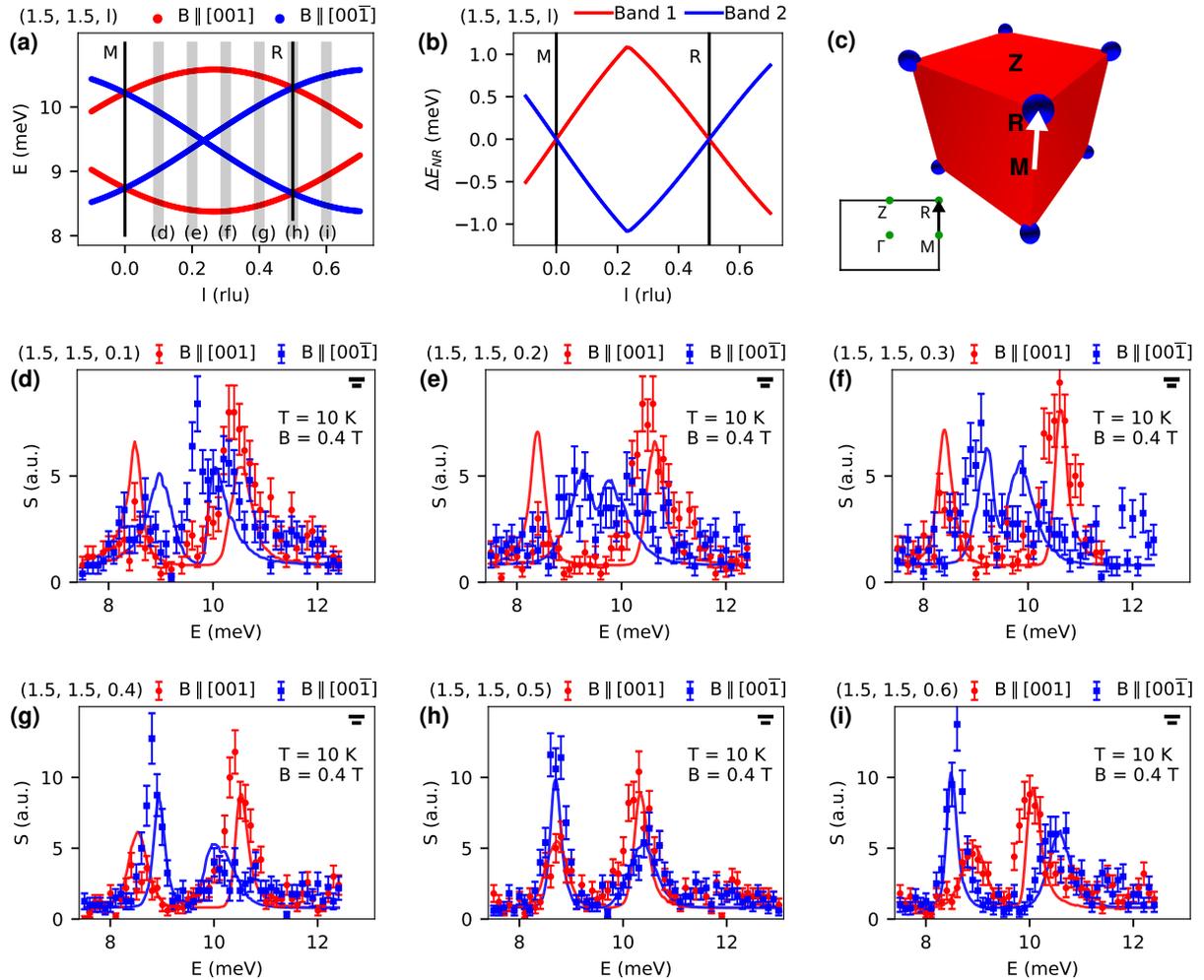

For our linear spin-wave calculations we adapted the effective model developed by Luo \textit{et al.} \cite{luo2020low}.
In the model, the 16 magnetic Cu sites in the \cso{} unit cell form four tetrahedral clusters of four Cu atoms each.
The exchange between the local moments on the clusters of four Cu ions is so strong
that the field-polarized phase is effectively described by a ferromagnetic configuration of
spin-one moments on a slightly distorted fcc lattice.
The effective Heisenberg Hamiltonian describing these effective spin-one moments is given by
\begin{align} \label{eq:hamiltonian}
	H &= \sum_{\alpha=1,2}\sum_{\langle ij\rangle_{\alpha}}
	J_{\alpha}\hat{\mathbf{S}}_i \cdot \hat{\mathbf{S}}_j +
		\mathbf{D}_{\alpha}\cdot \left(\hat{\mathbf{S}}_i \times \hat{\mathbf{S}}_j \right) \nonumber \\
	& + \sum_{i} g_e \mu_B \mathbf{B} \cdot \mathbf{S}_i,
\end{align}
with $J_{\alpha}$ and $\mathbf{D}_{\alpha}$ the Heisenberg and Dzyaloshinskii-Moriya interactions (DMI)
on the two symmetrically non-equivalent nearest-neighbor exchange paths $\alpha=1, 2$, respectively \cite{luo2020low}.
Luo \textit{et al.} originally introduced DMI vectors to account for the inability of previous models
to reproduce zone-boundary effects like the splitting of the modes near the corner, $R$ \cite{luo2020low}.
For $J_{\alpha} < 0$ and small $\mathbf{D}_{\alpha}$, the classical ground state is ferromagnetic.
The last term in Eq. (\ref{eq:hamiltonian}) is the Zeeman energy, with the external magnetic field $\mathbf{B}$,
and the electron g-factor $g_e \approx 2$.

Treatment of experimental data was performed using the \textit{Takin} \cite{Takin2023, takin} software suite
and a bespoke spin-wave code \cite{magpie}.
The software models magnetic systems and performs linear spin-wave calculations,
which it convolutes with the instrumental resolution function in order
to simulate the observed scan data or to refine open parameters in the model.
The programs compute the linear spin-wave Hamiltonians and spin-spin correlation functions
via the general theory outlined by Toth and Lake \cite{Toth2015} and
the resolution function via the Popovici formalism \cite{Popovici1975}.
More details on the model and data treatment are given in the \suppl{}.

Fig. \ref{fig:B001} shows the $\left(1.5,\ 1.5,\ l \right)$ dispersion from the edge midpoint $M = \left(1.5,\ 1.5,\ 0 \right)$
of the nuclear Brillouin zone towards the corner point $R = \left(1.5,\ 1.5,\ 0.5 \right)$.
The magnetic field is oriented along $B_l = \left[ 001 \right]$ and $-B_l = \left[ 00\bar{1} \right]$, respectively.
Panel (a) of Fig. \ref{fig:B001} depicts the theoretical dispersion branches,
panel (b) shows the difference in energy by which magnon creation is shifted from magnon annihilation,
$\Delta E_{NR} = \left|E \left(q \right)\right| - \left|E \left( -q \right)\right|$.
A maximum shift of energies of $\Delta E_{NR} \approx \pm 1\ \mathrm{meV}$ is therefore expected at a central position between
$M$ and $R$, at $l \approx 0.25\ \mathrm{rlu}$.
The measured data and the resolution-convoluted spectra are depicted in panels (d)-(i).
The dispersion differs dramatically for the two field directions with the branches changing from non-crossing to crossing
in the observed region between $M$ and $R$.
The inversion of the field effectively mirrors the dispersion branches at 0.5 rlu in the $l$ component.
When comparing the $\left(1.5,\ 1.5,\ 0.4 \right)$ and $\left(1.5,\ 1.5,\ 0.6 \right)$ scans (panels (g) and (i) in Fig. \ref{fig:B001}),
for example, the dispersion branches for the field along $\left[ 001 \right]$ trade places,
and flip back to their original energies when changing the field polarization from $\left[ 001 \right]$ to $\left[ 00\bar{1} \right]$.
The spectral weights of the modes, on the other hand, do not recover upon changing the field.

The solid lines in panels (d)-(i) of Fig. \ref{fig:B001}
were determined using a convolution fit between the linear spin-wave theory \cite{Toth2015}
and the instrumental resolution function \cite{Takin2023, Popovici1975}.
To that end, we set the magnetic interaction parameters, $J_{\alpha = 1}$ and $J_{\alpha = 2}$,
and the components of the DMI vectors,
$\mathbf{D}_{\alpha = 1}$ and $\mathbf{D}_{\alpha = 2}$,
of the model's two sub-lattices as open parameters and fitted them using all the relevant
scan data from \textit{Thales}.
Furthermore, a single global scaling parameter was fitted to normalize all spectral weights.
Table \ref{tab:constants} lists the fitted exchange constants and DMI vectors.
With the fitted parameters we obtain an excellent match between theory and experiment.
The imperfect correspondence that is uniquely located at low $l$ is due to the simplifying effective
model and the simultaneous fitting to all data.
The model parameters where the regions of our scans have most influence are in general agreement
with previously published values \cite{luo2020low},
these are $J_{\alpha = 1,2}$ and the $z$ components of $\mathbf{D}_{\alpha = 1,2}$.
They are crucial for obtaining the correct magnon energies.
While the influence of the $x$ and $y$ components of the DMI vectors on the fitted energies is limited,
as is their predictive value for the present case, they do have a slight effect
on the spin-spin correlation function and thus the peak intensities.
The effective model reproduces the correct qualitative picture in all cases.

\begin{table}[h]
\begin{tabular}{c|c|c|c}
    Param.    & Here     & Influence    & Luo \textit{et al.} \cite{luo2020low} \tabularnewline
    \hline
    \hline

    \hline
    $J_{\alpha = 1}$      & $-0.601 \pm 0.0067$   & high       & -0.58 \tabularnewline
    \hline
    $J_{\alpha = 2}$      & $-0.971 \pm 0.0078$   & high       & -0.93 \tabularnewline
    \hline

    \hline
    $D_{\alpha = 1, x}$   & $0.618 \pm 0.048$   & low-medium  & 0.24 \tabularnewline
    \hline
    $D_{\alpha = 1, y}$   & $0.018 \pm 0.059$   & low         & -0.05 \tabularnewline
    \hline
    $D_{\alpha = 1, z}$   & $0.107 \pm 0.015$   & high        & -0.15 \tabularnewline
    \hline

    \hline
    $D_{\alpha = 2, x}$  & $0.755 \pm 0.047$   & low-medium  & -0.16 \tabularnewline
    \hline
    $D_{\alpha = 2, y}$  & $0.234 \pm 0.061$   & low         & -0.1 \tabularnewline
    \hline
    $D_{\alpha = 2, z}$  & $0.303 \pm 0.014$   & high        & 0.36 \tabularnewline
\end{tabular}
\caption{Fitted model parameters compared to previously published results \cite{luo2020low}.
The \textit{influence} column indicates how strong each parameter influences the spectrum in the scanned regions
at the zone boundary.}
\label{tab:constants}
\end{table}

We observe strong NR magnon effects along the $l$ direction of the Brillouin zone boundary of \cso{},
resulting in energy shifts of up to $\Delta E_{NR} \approx 1\ \mathrm{meV}$.
The effects manifest themselves upon changing the polarity of the applied magnetic field along the $l$ direction.
Inverting the field mirrors the magnetic band crossing located between the $M$ and the $R$ symmetry
points of the Brillouin zone towards $M$ and $-R$.
In its paramagnetic phase, the combination of time-reversal symmetry together with any of the three two-fold screw
rotations along the principal crystal axes is a symmetry of the system. Ferrimagnetic order along one of these axes leaves
the combined symmetry intact in directions perpendicular to the magnetic moment, but breaks it for directions that are parallel.
There, only the screw rotation remains intact without time-reversal (see the \suppl{}).
The aforementioned band crossing is a Weyl point \cite{zhang2020magnonic} that is enforced by the
two-fold screw axis along the $z$ direction together with broken time-reversal symmetry \cite{Alpin2023}.
The latter is caused by the stabilization of the ferrimagnetic order due to the application of a magnetic
field along that direction.

The strong NR magnetic dynamics is reflected by the large magnitudes of the two DMI vectors.
Their overall magnitudes are larger than in previous works \cite{luo2020low} that were performed
without taking field-dependent effects into account.
A caveat for the limit of the effective model is that our parameters are only fitted to our detail measurements
of the zone boundary magnons, whereas previous works performed fits in all directions at zero field.

The present results are unique in that a strong DMI normally entails a damping of the high-energy modes \cite{Chernyshev2016}
which can be modelled by fourth-order magnon processes.
However, in \cso{} we detect no damping between $M$ and $R$ beyond the instrumental resolution.
This can be explained by the forced collinear alignment of the spins in the direction of the applied field,
which eliminates the leading damping contributions coming from the DMI.
Other studies observed line broadening restricted to the $X$ point \cite{luo2020low}.

Previous low-$q$ studies in \cso{} \cite{Seki2020, Che2024} found NR shifts of
$\approx 0.3\ \mathrm{meV}$ for the counter-clockwise mode of the skrymion phase,
$\approx 0.2\ \mathrm{meV}$ for the field-polarized phase,
and $\lesssim 0.05\ \mathrm{meV}$ for the conical phase near the $\Gamma$ point.
Phenomenologically similar NR effects on magnons are observed in the low-energy modes in the
skyrmion \cite{Weber2022skx}, conical \cite{Weber2019coni}, and field-polarized \cite{Sato2016} phases of MnSi.
In the helical and conical phases of both \cso{} and MnSi, the NR shift in energy is very low
inside the first magnetic Brillouin zone ($\lesssim 0.01\ \mathrm{meV}$),
instead the effect manifests itself in shifts of the spectral weight between magnon creation and annihilation.

We performed several tests to exclude other explanations for the observed effects.
Above the critical temperature, all signals vanish, proving their magnetic origin.
The magnetic field polarity was switched by inverting the direction of the current through the coils,
not by turning the magnets by $180^{\circ}$, thereby ruling out any change in neutron absorption
patterns caused by the sample environment.
Finally, the field direction was only changed above the critical temperature,
eliminating any hysteresis effects.

In summary, using inelastic neutron scattering covering both high energy transfers and
high resolution, we observe strong NR effects in the magnon excitation
spectrum of \cso{} at the maximum possible reduced momenta at the boundary of the nuclear Brillouin zone.
The NR response is strongest at a central momentum between the $M$ and the $R$ point.
Even though the effect is driven by the DMI, the magnons possess minimal damping,
which makes the system especially interesting for magnonic applications.

\textit{Acknowledgements.} We wish to thank Helmuth Berger,
formerly at the \'Ecole Polytechnique F\'ed\'erale de Lausanne in Switzerland
and now retired, for growing and providing the \cso{} crystal.
Thanks to Emmanuel Villard for technical support at \textit{Thales}.
We want to thank Marc Wilde, Markus Garst, and Christian Back for valuable discussions.

T.W. designed the experiments and performed data analysis.
T.W. and M.S. performed the experiments.
N.H. and T.W. each implemented the model.
T.W. wrote the manuscript with input from N.H. and C.P.
P.S. supported the experiments at the \textit{Thales} spectrometer.
C.P. initiated the project.
A.P.S. supervised the theoretical work.
All authors discussed the results and reviewed the manuscript.

The experiments were performed at the triple-axis spectrometer \textit{Thales} \cite{Thales}
at the Institut Laue-Langevin under proposal 4-01-1820 (DOI: \href{https://doi.ill.fr/10.5291/ILL-DATA.4-01-1820}{10.5291/ILL-DATA.4-01-1820}).
The source code and data is available under DOI \href{https://doi.org/10.5281/zenodo.15006081}{10.5281/zenodo.15006081} \cite{data}.

This study was funded by the Deutsche Forschungsgemeinschaft (DFG, German Research Foundation) under TRR360 (Constrained Quantum Matter, Project No. 492547816), SPP2137 (Skyrmionics, Project No. 403191981, Grant PF393/19), and the excellence cluster MCQST under Germany's Excellence Strategy EXC-2111 (Project No. 390814868). Financial support by the European Research Council (ERC) through Advanced Grant No. 788031 (ExQuiSid) is gratefully acknowledged.

%

\end{document}


\newcommand{\cso}{Cu$_{2}$OSeO$_{3}$}

\title{Supplementary Materials for ``Non-Reciprocal Zone Boundary Magnon Propagation in \cso{}''}

\newcommand{\ill}{Institut Laue-Langevin (ILL), 71 avenue des Martyrs, 38000 Grenoble, France}
\newcommand{\tum}{Physik-Department, Technische Universit\"at M\"unchen (TUM), James-Franck-Str. 1, 85748 Garching, Germany}
\newcommand{\mlz}{Heinz-Maier-Leibnitz-Zentrum (MLZ), Technische Universit\"at M\"unchen (TUM), Lichtenbergstr. 1, 85747 Garching, Germany}
\newcommand{\frm}{Forschungsneutronenquelle Heinz-Maier-Leibnitz (FRM-II), Lichtenbergstr. 1, 85747 Garching, Germany}
\newcommand{\epfl}{\'Ecole Polytechnique F\'ed\'erale de Lausanne (EPFL), CH-1015 Lausanne, Switzerland}
\newcommand{\fkf}{Max-Planck-Institut f\"ur Festk\"orperforschung, Heisenbergstr. 1, D-70569 Stuttgart, Germany}
\newcommand{\jcns}{J\"ulich Centre for Neutron Science (JCNS), Lichtenbergstr. 1, 85748 Garching, Germany}
\newcommand{\mcqst}{Munich Center for Quantum Science and Technology (MCQST), Schellingstr. 4, 80799 Munich, Germany}
\newcommand{\zqe}{Zentrum f\"ur QuantumEngineering (ZQE), Am Coulombwall 3a, 85748 Garching, Germany}

\author{T. Weber}
\email[Corresponding author: ]{tobias.weber@ill.fr}
\affiliation{\ill}

\author{N. Heinsdorf}
\affiliation{\fkf}

\author{M. Stekiel}
\affiliation{\tum}
\affiliation{\jcns}

\author{P. Steffens}
\affiliation{\ill}


\author{A. P. Schnyder}
\affiliation{\fkf}

\author{C. Pfleiderer}
\affiliation{\tum}
\affiliation{\frm}
\affiliation{\mcqst}
\affiliation{\zqe}

\date{\today}

\begin{abstract}
	These Supplementary Materials describe the spin-wave theory as well as the experimental
	and data treatment techniques that we used for our study.
\end{abstract}

\maketitle

\section{Magnetic Model}
\label{sec:model}
\cso{} crystallizes in the cubic $\mathrm{P2_13}$ space group and has a lattice constant of $a = 8.92\ \textup{\AA}$.
Of interest for the spin-waves are the magnetically active $\mathrm{Cu^{2+}}$ atoms,
of which one unit cell houses four clusters of four tetrahedrally organized atoms each,
see Fig. \ref{fig:model} (a).
Three of the tetrahedral $\mathrm{Cu^{2+}}$ atoms couple ferromagnetically to one another
and antiferromagnetically to the fourth $\mathrm{Cu^{2+}}$ atom \cite{luo2020low}.

The effective magnetic model devised by Luo \textit{et al.} \cite{luo2020low}
approximates each of the four tetrahedra as one single magnetic site having spin $1$,
pointing in the direction of the originally ferromagnetically coupled $\mathrm{Cu^{2+}}$ atoms
or -- in the field-polarized phase as in the present case -- along an applied external field.
The positions of the four effective magnetic sites are given in Tab. \ref{tab:sites}.
The two nearest possible exchange paths,
which each couple sites at a distance of $d = 6.286\ \textup{\AA}$ and are both ferromagnetic,
are shown in Tab. \ref{tab:terms} and Fig. \ref{fig:model} (b) and (c).
Both of them include Heisenberg and Dzyaloshinskii-Moriya interactions.

\begin{table}[h]
\begin{tabular}{|c|c|}
	\hline
	Site Index  & Position            \tabularnewline
	\hline
	\hline
	1           & $\left(\frac{1}{4}\frac{1}{4}\frac{1}{4}\right)$  \tabularnewline
	\hline
	2           & $\left(\frac{3}{4}\frac{1}{4}\frac{3}{4}\right)$  \tabularnewline
	\hline
	3           & $\left(\frac{1}{4}\frac{3}{4}\frac{3}{4}\right)$  \tabularnewline
	\hline
	4           & $\left(\frac{3}{4}\frac{3}{4}\frac{1}{4}\right)$  \tabularnewline
	\hline
\end{tabular}
\caption{Magnetic sites of the effective model. All four positions are symmetry-equivalent in the $\mathrm{P2_13}$ space group.
	The sites are plotted as red spheres in Fig. \ref{fig:model} (b), (c).}
\label{tab:sites}
\end{table}

\begin{table}[h]
\begin{tabular}{|c|c|c|c|}
	\hline
	Exchange Index  & Sites         & Super-Cell          & Interactions     \tabularnewline
	\hline
	\hline
	1           & $2 \rightarrow 1$  & $\left[ 100 \right]$  & $J_{\alpha=1}$ and $\bm{D}_{\alpha=1}$  \tabularnewline
	\hline
	2           & $2 \rightarrow 1$  & $\left[ 001 \right]$  & $J_{\alpha=2}$ and $\bm{D}_{\alpha=2}$  \tabularnewline
	\hline
\end{tabular}
\caption{Reference magnetic exchange paths. Each of the two reference exchange paths corresponds
	to twelve symmetry-equivalent paths in the $\mathrm{P2_13}$ space group.
	The exchange paths are plotted as green (index 1) and turquois (index 2) lines in Fig. \ref{fig:model} (b), (c).}
\label{tab:terms}
\end{table}

\subsection{Symmetry Arguments}
The availability of non-reciprocal responses can be deduced from a symmetry analysis using the system's space group.
Time-reversal symmetry (TRS) acts on the dynamical structure factor as $\theta: \bm{S}_{\bm{q}} \rightarrow \mathbf{S}_{-\bm{q}}$
and is broken by magnetic order, which, in \cso{}, can be enforced by an external magnetic field.
Here, $\bm{S}$ designates the spin, and $\bm{q}$ the momentum transfer
If the lattice and spin degrees of freedom are decoupled, i.e. in the absence of spin-orbit interactions (SOI),
the composition of TRS with a unitary spin rotation remains a symmetry of the system,
leading to a reciprocal magnon dispersion, $\bm{S}_{\bm{q}} = \bm{S}_{-\bm{q}}$.
SOI is constrained by inversion symmetry.
Therefore, if inversion symmetry is broken, SOI couples the spin and lattice degrees of freedom
and the magnon dispersion can become non-reciprocal, that is,  $\bm{S}_{\bm{q}} \neq \bm{S}_{-\bm{q}}$.
Additional crystallographic symmetries can further restrict $\bm{S}_{\bm{q}}$.
In the paramagnetic phase of \cso{}, twofold screw axes about the three principal axes (together with spinless TRS)
enforce a reciprocal dispersion relation along all momentum directions \cite{ortho}.
In the field-polarized phase, the combination of TRS and screw rotations that are not perpendicular to the field are broken,
and thus the magnon dispersion becomes non-reciprocal along that direction.
The screw rotation about the field direction without TRS remains a symmetry and can enforce degeneracies
in the spectrum \cite{Wilde2021} along that direction.

\subsection{Magnetic Phases}
While our work exclusively concerns the field-polarized and effectively ferromagnetic phase,
Tab \ref{tab:phases} gives a short overview of all the magnetic phases in \cso{}.
A comprehensive study can be found in, e.g., Ref \onlinecite{Qian2018}.

Increasing the magnetic field starting from the helically ordered phase cants the spins towards
the field direction forming the conical phase, until -- in the field-polarized phase --
they point entirely along the field direction.
At the border of the conical phase a skyrmion order can be found consisting of topologically
protected spin vortices.

\begin{table}[h]
\begin{tabular}{|c|c|c|}
	\hline
	Phase               & Temperature                      & Field             \tabularnewline
	\hline
	\hline
	Helical              & $0$ - $59\ \mathrm{K}$     & $\approx 0 - 15\ \mathrm{mT} $ \tabularnewline
	\hline
	Conical             & $0$ - $59\ \mathrm{K}$     & $\approx 15 - 45\ \mathrm{mT}$ \tabularnewline
	\hline
	Skyrmion          & $\approx 58 \ \mathrm{K}$ & $\approx 10\ \mathrm{mT}       $ \tabularnewline
	\hline
	Field-polarized  & $0$ - $59\ \mathrm{K}$     & $> 45 \ \mathrm{mT}                $ \tabularnewline
	\hline
	Paramagnetic   & $> 59\ \mathrm{K}$            & -- \tabularnewline
	\hline
\end{tabular}
\caption{Magnetic phases in \cso{} for a field along $\left[001\right]$.
	All values are approximated from the diagrams in Ref. \onlinecite{Qian2018}.}
\label{tab:phases}
\end{table}

\subsection{Implementation of the Model}
For our work we exclusively used our in-house software \cite{Takin2023, Magpie} for the
implementation of the magnetic model and the calculation of the magnon dispersion relations.
We include all source code used for the present paper in Ref. \onlinecite{data}.
Furthermore, we also provide an alternate but equivalent implementation
of the model in \textit{Sunny} \cite{Sunny} for easy verification, see Listing \ref{lst:sunnymodel}.

{
\captionsetup[figure]{name=Listing}
\begin{figure}[htb]
	\begin{lstlisting}[basicstyle=\tiny, tabsize=4]
using Sunny, GLMakie, Printf

# interaction constants
J1 = -0.601
J2 = -0.971
D1 = [ 0.618, 0.018, 0.107 ]
D2 = [ 0.755, 0.234, 0.303 ]

@printf("Setting up magnetic sites...\n")
magsites = Crystal(lattice_vectors(
	8.89, 8.89, 8.89, 90, 90, 90),
	[ [ 0, 0, 0 ] ], 198)

# spin magnitudes and magnetic system
magsys = System(magsites,
	[ 1 => Moment(s = 1, g = 2.002) ], :dipole)

# set spin directions
polarize_spins!(magsys, [ 0, 0, -1 ])
@printf("%s", magsites)

@printf("Setting up magnetic couplings...\n")
set_exchange!(magsys,
	[
		  J1     D1[3]  -D1[2];
		-D1[3]      J1   D1[1];
		 D1[2]  -D1[1]      J1
	], Bond(3, 1, [ 1, 0, 0 ]))
set_exchange!(magsys,
	[
		  J2     D2[3]  -D2[2];
		-D2[3]      J2   D2[1];
		 D2[2]  -D2[1]      J2
	], Bond(3, 1, [ 0, 0, 1 ]))

# external field
phys_units = Units(:meV, :angstrom)
set_field!(magsys, -[ 0, 0, -1 ] * 0.4 * phys_units.T)

@printf("Plotting magnetic structure...\n%s\n", magsys)
view_crystal(magsys, refbonds = 2)
plot_spins(magsys)

@printf("Calculating spin-waves...\n")
calc = SpinWaveTheory(magsys; measure = ssf_perp(magsys))
bands = intensities_bands(calc,
	q_space_path(magsys.crystal,
	[[ 1.5, 1.5, -0.5 ], [ 1.5, 1.5, 0.5 ]], 512))
plot_intensities(bands; units = phys_units)
	\end{lstlisting}
	\caption{\textit{Sunny} \cite{Sunny} code for the magnetic model.}
	\label{lst:sunnymodel}
\end{figure}

\addtocounter{figure}{-1}
}

For our study we applied an external magnetic field along $c$.
In the field-polarized phase, the spins of the magnetic sites align towards the externally enforced direction.
Since the model with its original parameters \cite{luo2020low} was not able to reproduce our
detailed scans at the zone boundary, we fitted its parameters to our measured data,
while keeping the structure of the model, see Sec. \ref{sec:data_treatment}.

\begin{figure*}[htb]
\begin{centering}
\begin{tikzpicture}
	\draw (0, 0) node [ inner sep = 0] { \includegraphics[width=0.24\textwidth]{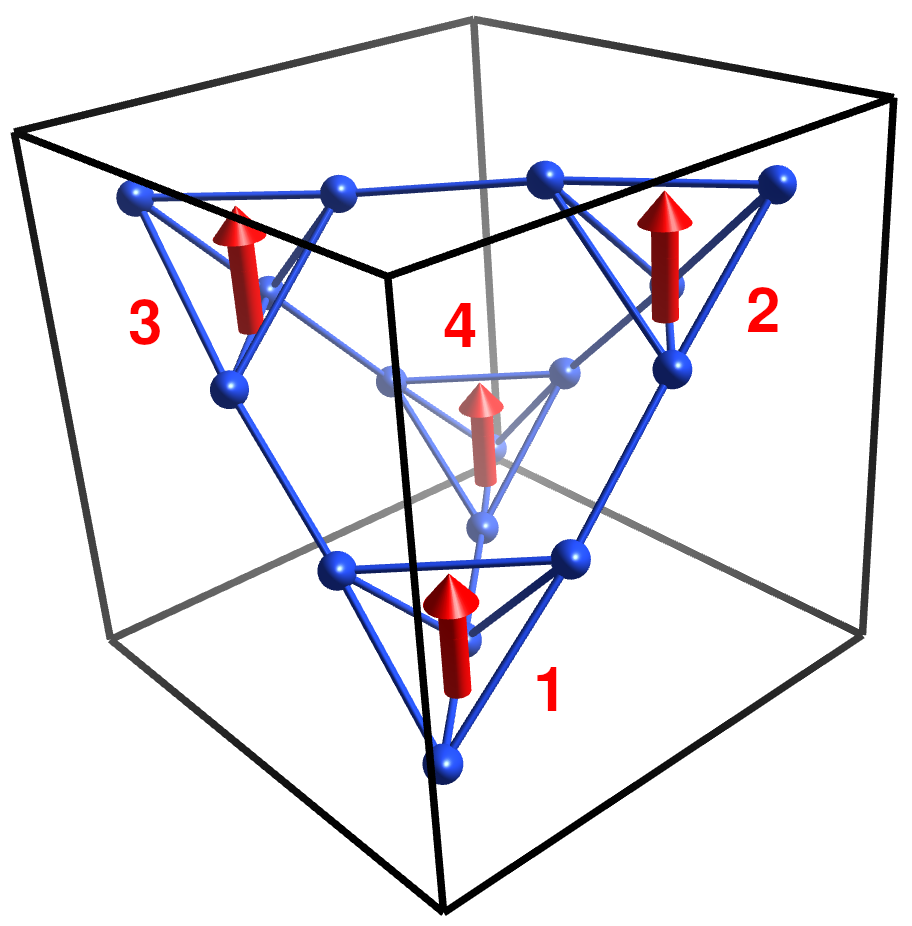} };
	\draw (-2, 2.5) node { \bf(a) };
\end{tikzpicture}
\hspace{0.5cm}
\begin{tikzpicture}
	\draw (0, 0) node [ inner sep = 0 ] { \includegraphics[width=0.3\textwidth]{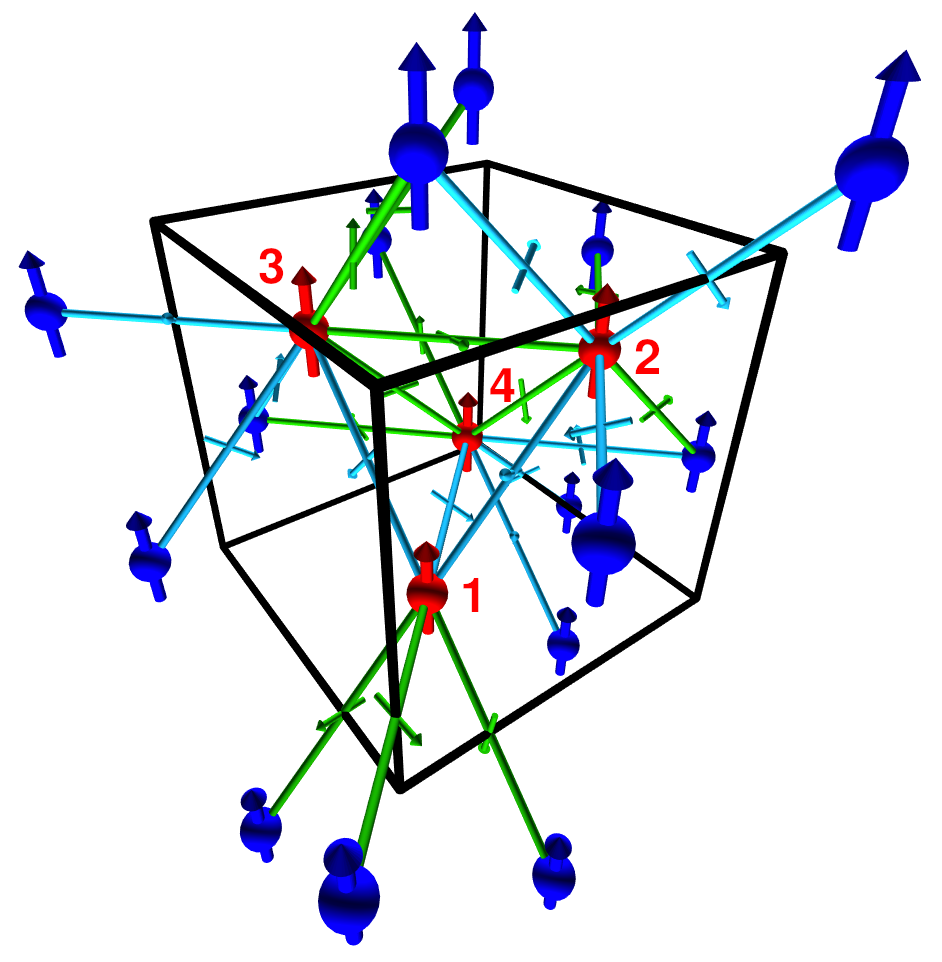} };
	\draw (-2, 2) node { \bf(b) };
\end{tikzpicture}
\hspace{0.5cm}
\begin{tikzpicture}
	\draw (0, 0) node [ inner sep = 0 ] { \includegraphics[width=0.27\textwidth]{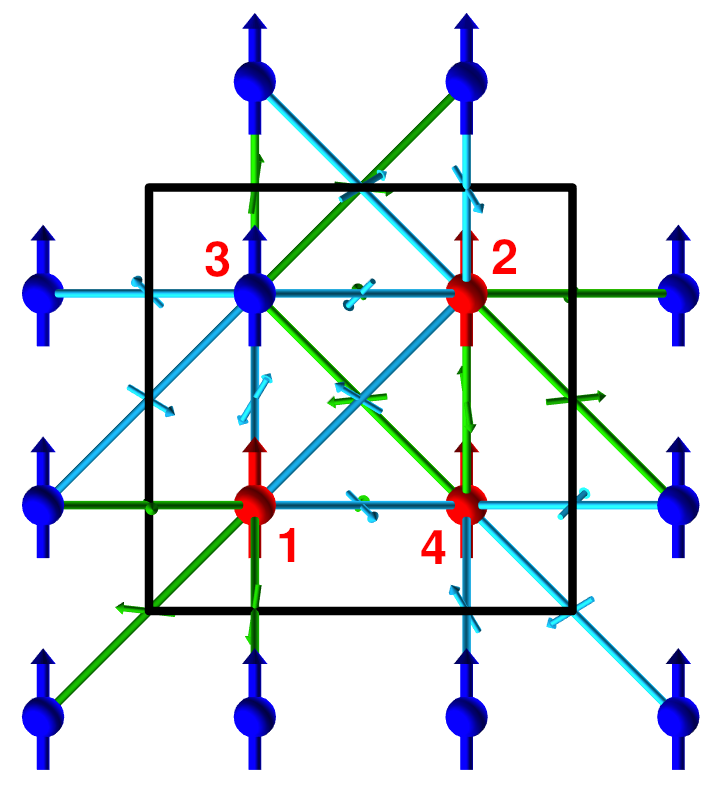} };
	\draw (-2, 2) node { \bf(c) };
\end{tikzpicture}
	\caption{Magnetic model.
	Panel (a): Four $\mathrm{Cu^{2+}}$ tetrahedra are each replaced by a single magnetic site in the
	middle of the corresponding tetrahedron. As we apply a magnetic field along the $z$ direction,
	the spins of the magnetic sites are polarized in that direction.
	Panels (b) and (c): The effective magnetic sites are shown as spheres, which are marked as red for sites
	within the first unit cell and blue for super-cell sites. The two possible exchange pathways are marked
	as green and turquois lines, which each have twelve symmetry-equivalent directions.
	The small arrows on the exchange paths show the direction of the Dzyaloshinskii-Moriya vectors.
	Panel (b) shows a perspective projection, (c) shows an orthographic view along the $b$ axis, with $a$ and $c$ in the drawing plane.
	We used the software \textit{Vesta} \cite{Vesta} for creating panel (a), and \textit{Magpie} \cite{Magpie} for panels (b) and (c).}
	\label{fig:model}
\end{centering}
\end{figure*}

\section{Experiment}
\subsection{Crystal}

\begin{figure}[htb]
\begin{centering}
\begin{tikzpicture}
	\draw (0, 0) node [ inner sep = 0] { \includegraphics[width=0.25\textwidth]{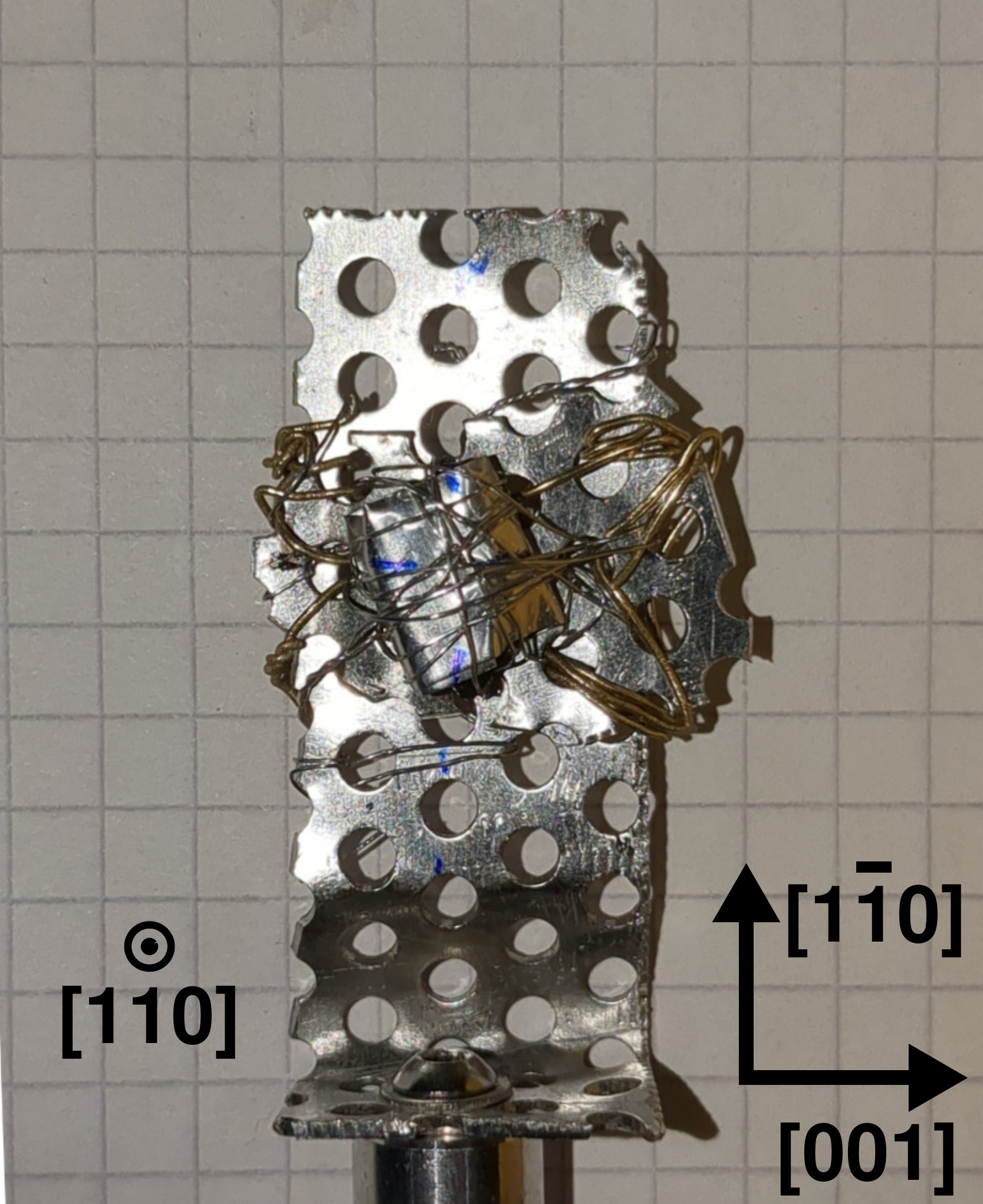} };
\end{tikzpicture}
	\caption{The \cso{} crystal (center) on its sample holder, aligned in the $\left[hhl \right]$ scattering plane
		and $\left[1\overline{1}0\right]$ pointing upwards.}
	\label{fig:xtal}
\end{centering}
\end{figure}

Our \cso{} crystal is shown in Fig. \ref{fig:xtal}, where it is mounted onto its sample holder, fixed in place with aluminum wires
and placed inside an aluminum pouch.
In this configuration, the crystal is aligned in the $\left[hhl \right]$ scattering plane.
The alignment was performed using neutron Laue diffraction at the \textit{OrientExpress} instrument \cite{OrientExpress} at the ILL.

\subsection{Data Treatment}
\label{sec:data_treatment}
Treatment of experimental data was performed using the \textit{Takin} \cite{Takin2023, takin} software suite
and its now separately available magnetic dynamics package, \textit{Magpie} \cite{Magpie}, both developed in-house.
Together, the programs allow to model magnetic systems and perform linear spin-wave calculations,
which are convoluted with the instrumental resolution function in order
to simulate the observed scan data or to refine open parameters in the model.
In \textit{Magpie}, the linear spin-waves are calculated via the general theory outlined by Toth and Lake \cite{Toth2015}
and the magnetic model discussed in Sec. \ref{sec:model}.
The resolution function is calculated using the Popovici formalism \cite{Popovici1975}.

\paragraph{\bf{Resolution}}
Calculating the resolution of a neutron spectrometer is usually done in Gaussian approximation
of the transmission functions of the spectrometer's neutron-optical components.
The transmission function of the full instrument is transformed into the $\left(\bm{Q}, E\right)$ coordinate space of the sample and yields a four-dimensional
covariance matrix $C$ and a scalar normalization factor $R_0$, see Ref. \onlinecite{Popovici1975} for details.

\begin{figure}[h]
\begin{centering}
\begin{tikzpicture}
	\draw (0, 0) node [ inner sep = 0 ] { \includegraphics[width=0.49\textwidth]{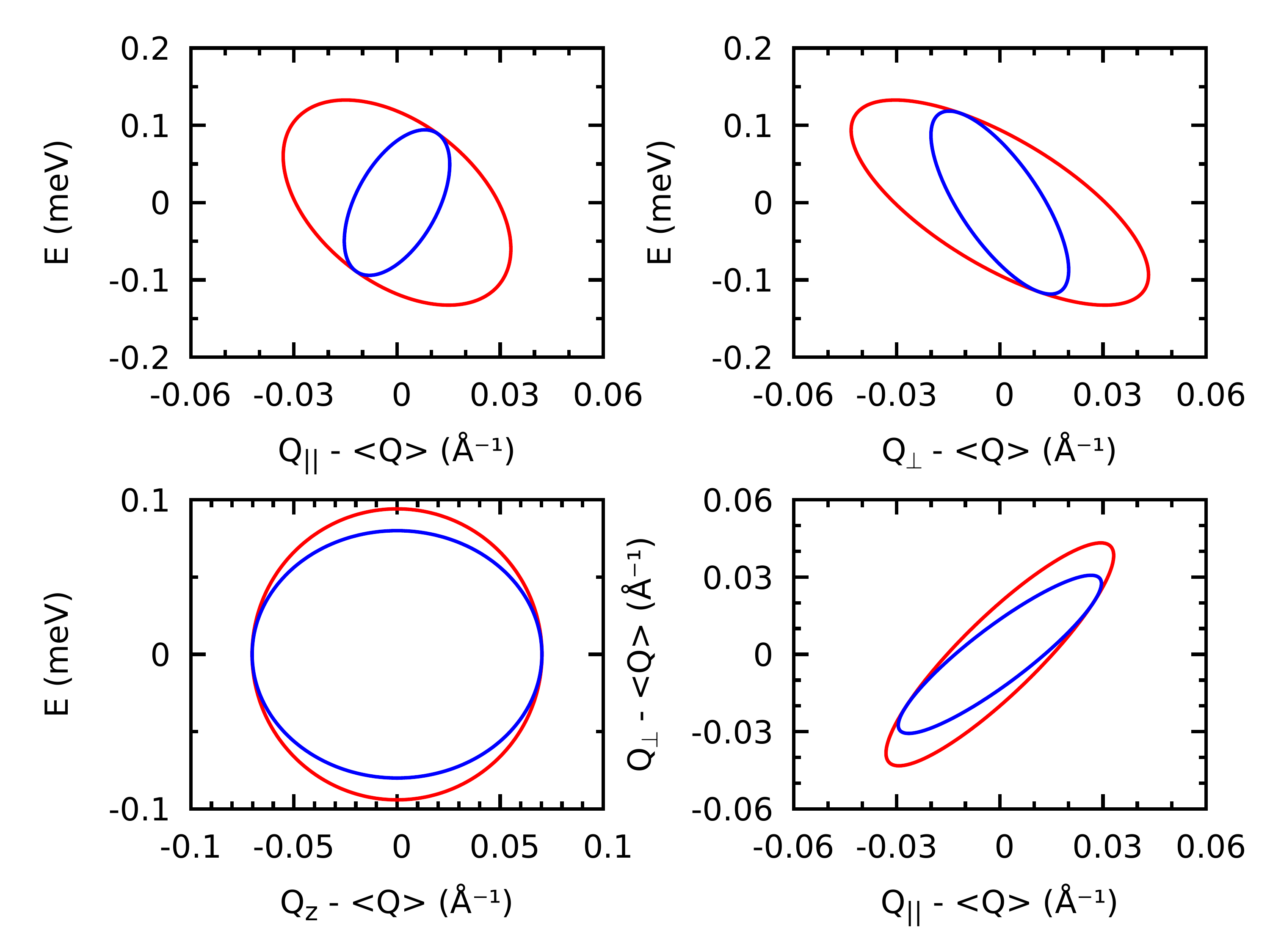} };
\end{tikzpicture}
	\caption{Resolution function for \textit{Thales} at $Q = \left(1.5,\  1.5,\  0.5 \right)$ and $E = 10 \ \mathrm{meV}$.
	The blue lines are slices through the four-dimensional resolution ellipsoid, the red lines are projections.}
	\label{fig:reso}
\end{centering}
\end{figure}

Fig. \ref{fig:reso} shows a typical resolution function, here at $Q\ =\ \left(1.5,\  1.5,\  0.5 \right)$ and $E = 10 \ \mathrm{meV}$,
for the \textit{Thales} \cite{THALES} spectrometer and our instrumental configuration.
At \textit{Thales} the incoherent FWHM energy resolution is 0.27 meV and the coherent one is 0.16 meV.

\paragraph{\bf{Resolution Convolution}}
Normally-distributed Monte-Carlo neutrons are generated with covariance $C$
and are used to probe the dynamical structure factor of the magnons, $S\left(\bm{Q},E\right)$.
The results of all simulated Monte-Carlo events are summed.
The whole process is summarized in the following integral \cite{Popovici1975}, where the
Gaussian distribution describing the instrumental resolution is given by the inner exponential factor:

\begin{dmath}
	I\left(\bm{Q}_{0},E_{0}\right)\ =\ R_{0} \cdot
		\intop d\bm{Q}dE \
			\exp\left(-\frac{1}{2}
				\left(\begin{array}{c} \bm{Q}-\bm{Q}_{0}\\ E-E_{0} \end{array}\right)^{T}
				\cdot C^{-1}\cdot
				\left(\begin{array}{c} \bm{Q}-\bm{Q}_{0}\\ E-E_{0} \end{array}\right)\right)
			\cdot S\left(\bm{Q},E\right).
	\label{eq:convo}
\end{dmath}

In Eq. \ref{eq:convo}, $\bm{Q}_0$ and $E_0$ name the nominal momentum and energy transfers for which the measurement takes place,
while $\bm{Q}$ and $E$ are the actual momentum and energy transfers over which we integrate via Monte-Carlo integration.
The inverse of the covariance matrix, $C^{-1}$, is usually called the resolution matrix. $R_0$ is the normalization factor.
To yield a full spectrum, this integral is solved for every point of the scanned data set.

The intensity of the measured data is on an arbitrary scale, namely neutron counts for a given time span normalized to a radiation monitor
which is placed on the $k_i$ axis and which decouples the intensity from the monochromator efficiency.
As such, the neutron events are not on the same scale as the theoretical dynamical structure factor $S\left(\bm{Q}, E \right)$,
and we have to introduce a global intensity scaling parameter.
This parameter was determined for one scan and fixed to be the same for all other scans.

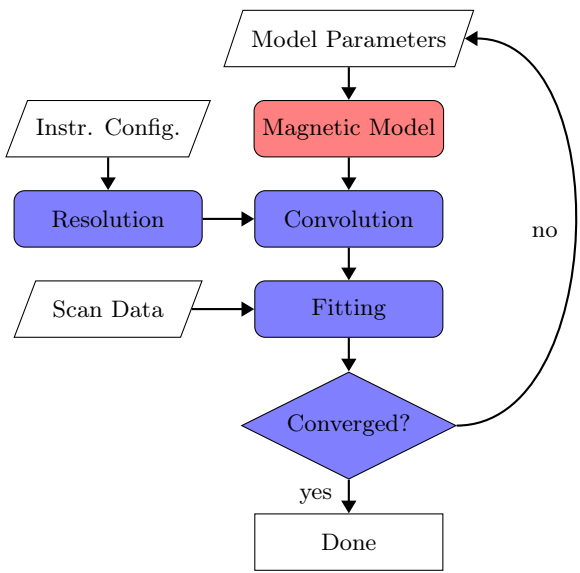
\begin{figure}[h!]
\begin{centering}
	\begin{tikzpicture}[node distance = 1.2cm]
		\tikzstyle{normal} = [rectangle,
			minimum width = 2.5cm, minimum height = 0.75cm,
			draw = black, fill = white, text centered, ]
		\tikzstyle{magpie} = [rectangle, rounded corners,
			minimum width = 2.5cm, minimum height = 0.75cm,
			draw = black, fill = red!50, text centered, ]
		\tikzstyle{takin} = [rectangle, rounded corners,
			minimum width = 2.5cm, minimum height = 0.75cm,
			draw = black, fill = blue!50, text centered, ]
		\tikzstyle{input} = [trapezium, trapezium stretches = true,
			minimum width = 2.5cm,  minimum height = 0.75cm,
			trapezium left angle = 60, trapezium right angle = 120,
			draw = black, fill = white, text centered, ]
		\tikzstyle{cond} = [diamond, aspect = 2,
			minimum width = 2.5cm, minimum height = 0.75cm,
			draw = black, fill = blue!50, text centered, ]

		\node(params)[input]{Model Parameters};
		\node(model)[magpie, below of = params]{Magnetic Model};
		\node(convo)[takin, below of = model]{Convolution};
		\node(reso)[takin, left of = convo, xshift = -2cm]{Resolution};
		\node(fitting)[takin, below of = convo]{Fitting};
		\node(data)[input, left of = fitting, xshift = -2cm]{Scan Data};
		\node(instr)[input, above of = reso]{Instr. Config.};
		\node(if_converged)[cond, below of = fitting, yshift = -0.35cm]{Converged?};
		\node(done)[normal, below of = if_converged, yshift = -0.35cm]{Done};

		\draw[thick, -Triangle] (params) to (model);
		\draw[thick, -Triangle] (model) to (convo);
		\draw[thick, -Triangle] (reso) to (convo);
		\draw[thick, -Triangle] (convo) to (fitting);
		\draw[thick, -Triangle] (data) to (fitting);
		\draw[thick, -Triangle] (instr) to (reso);
		\draw[thick, -Triangle] (fitting) to (if_converged);
		\draw[thick, -Triangle] (if_converged) to [out = 0, in = 0] node[anchor = east, xshift = -0.1cm]{no}(params);
		\draw[thick, -Triangle] (if_converged) to node[anchor = east, xshift = -0.1cm]{yes}(done);
	\end{tikzpicture}

	\caption{Resolution-convolution fitting workflow for refining the parameters of the magnetic model.
		The red-shaded and blue-shaded rectangles correspond to tasks performed
		by the \textit{Magpie} and \textit{Takin} software tools, respectively.}
	\label{fig:workflow}
\end{centering}
\end{figure}

\paragraph{\bf{Resolution Convolution Fitting}}
We set the magnetic interaction parameters and the components of the Dzyaloshinskii-Moriya vectors of the model's two sub-lattices
as open parameters and determined them using a global resolution-convolution fit of all the scan data simultaneously.
The fitting process works by starting from user-given initial parameters of the magnetic model,
for which we used the values given in Ref. \onlinecite{luo2020low}.
The fitter randomly varies the parameters and performs a $\chi^2$ test between the calculated resolution-convoluted theory curves
and the expected outcome that is given by the experimental data.
If a new minimum in $\chi^2$ is found, it updates the parameters, recalculates the magnetic model and restarts the workflow.
This is done until a $\chi^2$ minimum is accepted.
The workflow is depicted schematically in Fig. \ref{fig:workflow}.

\begin{figure*}[htb]
\begin{centering}
	\begin{tikzpicture}
		\draw (0, 0) node [ inner sep = 0 ] { \includegraphics[width=0.4\textwidth]{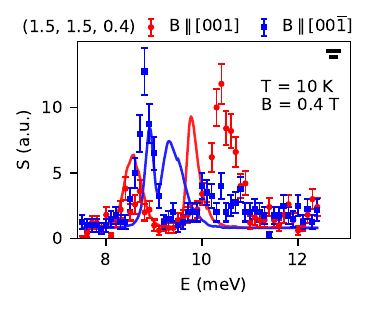} };
		\draw (0, 3) node { \bf \fontfamily{phv} \selectfont (a) Convolution using initial parameters };
	\end{tikzpicture}
	\hspace{0.75cm}
	\begin{tikzpicture}
		\draw (0, 0) node [ inner sep = 0, yshift = 3cm ] { \includegraphics[width=0.075\textwidth]{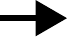} };
	\end{tikzpicture}
	\hspace{0.75cm}
	\begin{tikzpicture}
		\draw (0, 0) node [ inner sep = 0 ] { \includegraphics[width=0.4\textwidth]{figures/thales2_15_15_04_001.pdf} };
		\draw (0, 3) node { \bf \fontfamily{phv} \selectfont (b) Convolution using fitted parameters };
	\end{tikzpicture}

	\caption{Typical resolution-convolution results for (a) the initial unrefined model parameters and (b) the final fitted parameters.
	The initial parameters used in panel (a) correspond to the ones from Ref. \onlinecite{luo2020low} and are unable to reproduce our detail measurements.}
	\label{fig:before_after}
\end{centering}
\end{figure*}

\paragraph{\bf{Example Results}}
Fig. \ref{fig:before_after} shows the results for $\left(1.5,\ 1.5,\ 0.4 \right)$ both before (panel a)
and after (panel b) fitting the model's interaction parameters.
The initial parameters used in panel (a) correspond to the ones of Ref. \onlinecite{luo2020low}.
With its original parameters, the model does not correctly account for the energy and spin-spin correlation function
of the high-energy magnon branch, while the match with the low-energy branch is satisfactory.

Panel (b) of Fig. \ref{fig:before_after} is the same plot that is depicted in the main text.
Both the predicted magnon branches and the observed spectra match to a much higher degree.
It should also be noted that the fitted parameters were determined using all scans simultaneously,
not each single scan individually.
Such a procedure optimizes the model globally, but can lead to worse fits for individual scans than what
could be achieved by individual fits.

\subsection{Instrumental Configuration}
\label{sec:tas_config}

\begin{figure*}[htb]
\begin{centering}
\begin{tikzpicture}
	\draw (0, 0) node [ inner sep = 0 ] { \includegraphics[width=0.42\textwidth]{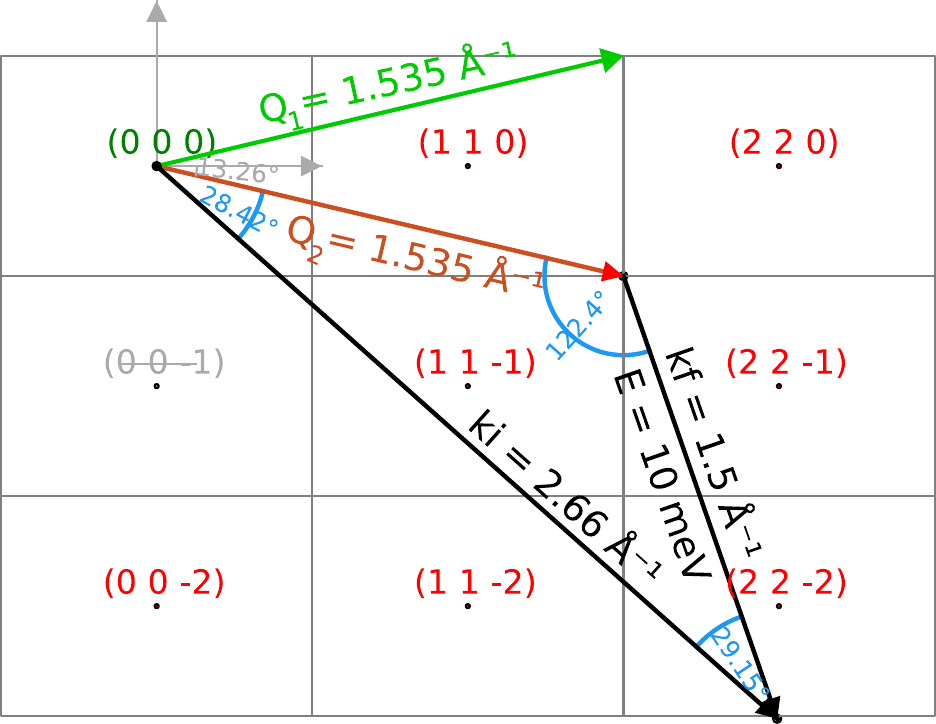} };
	\draw (-3.1, 3) node { \bf(a) };
\end{tikzpicture}
\begin{tikzpicture}
	\draw (0, 0) node [ inner sep = 0 ] { \includegraphics[width=0.25\textwidth]{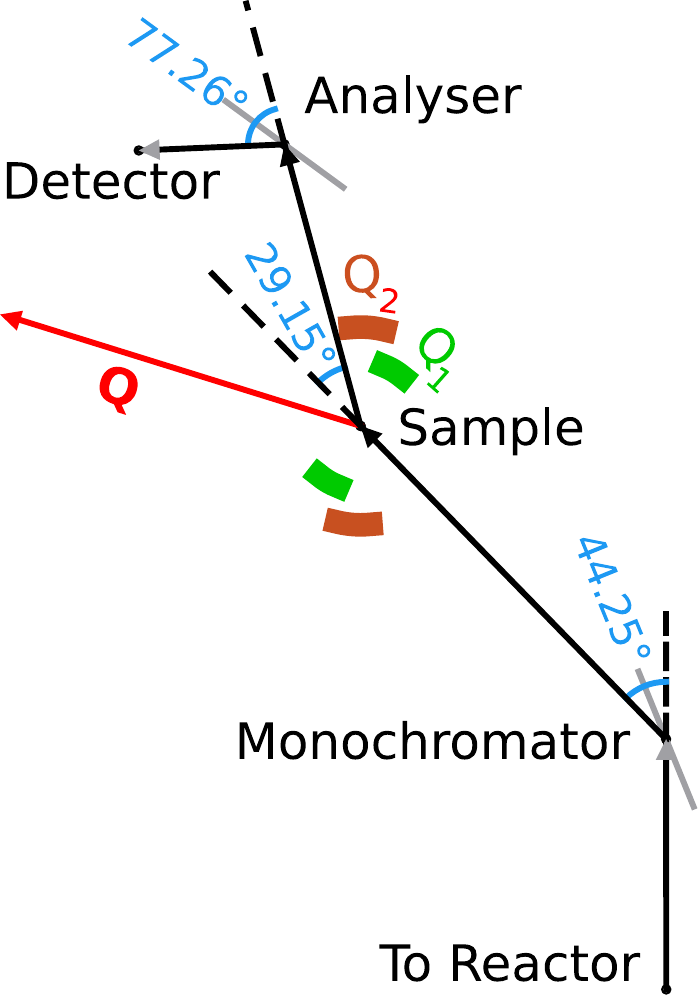} };
	\draw (-2, 2.75) node { \bf(b) };
\end{tikzpicture}
\begin{tikzpicture}
	\draw (0, 0) node [ inner sep = 0 ] { \includegraphics[width=0.25\textwidth]{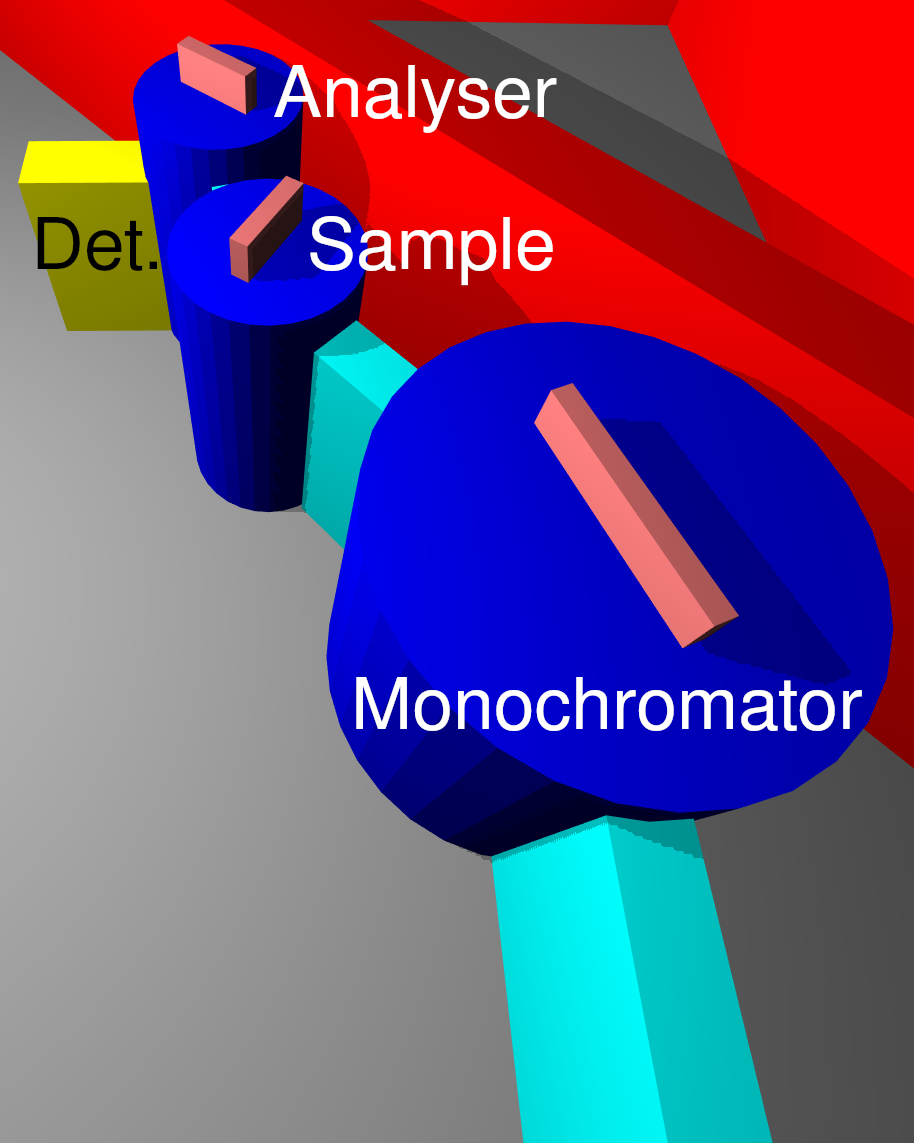} };
	\draw (-2, 3.1) node { \bf(c) };
\end{tikzpicture}
	\caption{Scattering geometry of our experiment at \textit{Thales}.
		(a) Two typical momentum transfers, namely $Q_1 = \left(1.5,\  1.5,\  +0.5 \right)$ and $Q_2 = \left(1.5,\  1.5,\  -0.5 \right)$, are shown.
		In addition, the scattering triangle corresponding to an energy transfer of $E = 10\ \mathrm{meV}$ is plotted for $Q_2$.
		(b) View of the instrumental angles for the given scattering triangle.
		The dark angles caused by the yokes of the magnet are marked
		as green circle segments around the sample position for $Q_1 = \left(1.5,\  1.5,\  +0.5 \right)$
		and as red circle segments for $Q_2 = \left(1.5,\  1.5,\  -0.5 \right)$.
		The position $Q_1$ is allowed, in $Q_2$ the yokes cut off parts of the scattered beam.
		(c) A schematic view of the instrumental configuration for $Q_{1,2}$ and $E$ shows that
		the spectrometer drives very close to its outer limits,
		which in the software \cite{TasPaths2023} is approximated by an effective wall (red).}
	\label{fig:tas_config}
\end{centering}
\end{figure*}

The setup of the triple-axis spectrometer \textit{Thales} \cite{THALES} used for our experiment is shown in Fig. \ref{fig:tas_config}.
The figure shows the scattering triangle (panel a) and the instrumental angles (panel b)
for a typical momentum transfer of $Q_{1,2} = \left(1.5,\  1.5,\  \pm 0.5 \right)$ and an energy transfer of $E = 10\ \mathrm{meV}$.
At $E = 10\ \mathrm{meV}$ we scatter from a relatively high incoming neutron wavenumber of $k_i = 2.66\, \textup{\AA}^{-1}$
down to a fixed $k_f = 1.5\, \textup{\AA}^{-1}$.
This way we profit both from the high resolution of a fixed $k_f$ in the cold energy region and from the relatively high flux
that \textit{Thales} still possesses even for thermal $k_i$.
Such a strong difference in $k_i$ and $k_f$ imposes some important restriction on the possible momentum and energy transfers.

The first restriction prevents the scattering triangle from being closed for energy transfers above $14.4\ \mathrm{meV}$
or below $\left(1.11,\  1.11,\  -0.5 \right)$, with the actual limit at the instrument already being below the theoretical limit of
energy transfer, and above the limiting momentum transfer.
This is due to the very low scattering angle in the vicinity of these limits.
It is so low that the direct beam from the monochromator would start to contaminate the signal.

The second restriction on scattering is given by the dark angles of the horizontal cryomagnet's yokes,
which are shown as green and red circle segments in Fig. \ref{fig:tas_config} (b), where green corresponds to the position $Q_1 = \left(1.5,\  1.5,\  +0.5 \right)$ and
red to $Q_2 = \left(1.5,\  1.5,\  -0.5 \right)$.
The cryomagnet has two dark angles of $\pm 10^{\circ}$ on opposing sides of the field direction \cite{MagnetH},
where the neutron beam is fully blocked, and penumbra regions which extend a bit further, where the beam is attenuated.
For $Q_1$ (green) the beam passes without problems, but for $Q_2$ (red) one of the dark angles is already so close to the $k_f$ axis that it fully cuts off parts of the magnon signals and attenuates the rest.

The third restriction is given by the maximum position the instrument can be driven to.
Here, we are severely limited due to very small monochromator angle moving the instrument close to the outer wall.
We calculated these limits by imposing a constraint on the instrument's available space in the software \textit{TAS-Paths} \cite{TasPaths2023}.
These constraints are shown as an effective wall marked in red in panel (c).

In total, possible scan positions on cold-neutron spectrometers like \textit{Thales} are strongly limited by several simultaneous constraints.
While the conditions would be more relaxed on a thermal spectrometer,
this would come at the cost of not being able to resolve the magnon bands or the non-reciprocal effects individually any more.

%